\documentstyle[11pt]{article}
\begin{document}

\title{{\bf Noether symmetry in $f(R)$ cosmology}}
\author{ Babak Vakili\thanks{%
email: b-vakili@sbu.ac.ir} \\
{\small Department of Physics, Shahid Beheshti University, Evin,
Tehran 19839, Iran}} \maketitle

\begin{abstract}
The Noether symmetry of a generic $f(R)$ cosmological model is
investigated by utilizing the behavior of the corresponding
Lagrangian under the infinitesimal generators of the desired
symmetry. We explicitly calculate the form of $f(R)$ for which such
symmetries exist. It is shown that the resulting form of $f(R)$
yields a power law expansion for the cosmological scale factor. We
also obtain the effective equation of state parameter for the
corresponding cosmology and show that our model can provide a
gravitational alternative to the quintessence.\vspace{5mm}\newline
PACS numbers: 04.20.Fy, 04.50.+h, 98.80.-k
\end{abstract}

\section{Introduction}
Observations of type Ia supernovae \cite{1} and cosmic microwave
background \cite{2} have revealed that the universe is undergoing
an accelerating phase in its expansion. The explanation of this
acceleration in the context of general relativity has stimulated a
myriad of ideas, the most notable of which is the introduction of
a mysterious cosmic fluid, the so-called dark energy \cite{3}. In
more recent times, modified theories of gravity, constructed by
adding correction terms to the usual Einstein-Hilbert action, have
opened a new window to study the accelerated expansion of the
universe where it has been shown that such correction terms could
give rise to accelerating solutions of the field equations without
having to invoke concepts such as dark energy \cite{4}.

In a more general setting, one can use a generic function, $f(R)$,
instead of the usual Ricci scalar $R$ as the action for the model.
Such $f(R)$ gravity theories have been extensively studied in the
literature over the past few years. In finding the dynamical
equations of motion one can vary the action with respect to the
metric (metric formalism), or view the metric and connections as
independent dynamical variables and vary the action with respect
to both independently (Palatini formalism) \cite{5}. In this
theory, the Palatini form of the action is shown to be equivalent
to a scalar-tensor type theory from which the scalar field
kinetic energy is absent. This is achieved by introducing a
conformal transformation in which the conformal factor is taken as
an auxiliary scalar field \cite{F}. As is well known, in the usual
Einstein-Hilbert action these two approaches give the same field
equations. However, in $f(R)$ gravity the Palatini formalism leads
to different dynamical equations due to nonlinear terms in the
action. There is also a third version of $f(R)$ gravity in which
the Lagrangian of the matter depends on the connections of the
metric (metric-affine formalism) \cite{6}.

In this letter we consider an ($n$+1)-dimensional flat FRW space-time
in the framework of the metric formalism of $f(R)$ gravity.
Following \cite{7}, we set up an effective Lagrangian in which the
scale factor $a$ and Ricci scalar $R$ play the role of independent
dynamical variables. This Lagrangian is constructed in such a way
that its variation with respect to $a$ and $R$ yields the correct
equations of motion as that of an action with a generic $f(R)$
mentioned above. The form of the function $f(R)$ appearing in the
modified action is then found by demanding that the Lagrangian
admits the desired Noether symmetry \cite{8}. By the Noether
symmetry of a given minisuperspace cosmological model we mean that
there exists a vector field $X$, as the infinitesimal generator of
the symmetry on the tangent space of the configuration space such
that the Lie derivative of the Lagrangian with respect to this
vector field vanishes. For a study of the Noether symmetry in
various cosmological models see \cite{9}. We shall see that by
demanding the Noether symmetry as a feature of the Lagrangian of
the model under consideration, we can obtain the explicit form of
the function $f(R)$. Since the existence of a symmetry results in
a constant of motion, we can integrate the field equations which
would then lead to a power law expansion of the universe.
\section{The phase space of the model}
In this section we consider a spatially flat FRW cosmology within
the framework of $f(R)$ gravity. Since our goal is to study models
which exhibit Noether symmetry, we do not include any matter
contribution in the action. Let us start from the
($n$+1)-dimensional action (we work in units where $c=16\pi G=1$)
\begin{equation}\label{A}
{\cal S}=\int d^{n+1}x\sqrt{-g}f(R),
\end{equation}
where $R$ is the scalar curvature and $f(R)$ is an arbitrary
function of $R$. By varying the above action with respect to
metric we obtain the equation of motion as
\begin{equation}\label{B}
\frac{1}{2}g_{\mu \nu}f(R)-R_{\mu
\nu}f'(R)+\nabla_{\mu}\nabla_{\nu}f'(R)-g_{\mu \nu}\Box f'(R)=0,
\end{equation}
where a prime represents differentiation with respect to $R$. We
assume that the geometry of space-time is described by the flat
FRW metric which seems to be consistent with the present
cosmological observations
\begin{equation}\label{C}
ds^2=-dt^2+a^2(t)\sum_{i=1}^{n}(dx^i)^2.
\end{equation}
With this background geometry the field equations read
\begin{equation}\label{D}
(n-1)\dot{H}+\frac{n(n-1)}{2}H^2=-\frac{1}{f'}\left[f'''\dot{R}^2+(n-1)H\dot{R}f''+f''\ddot{R}+\frac{1}{2}
(f-Rf')\right],\end{equation}
\begin{equation}\label{E}
H^2=\frac{1}{n(n-1)f'}\left[(f'R-f)-2n\dot{R}Hf''\right],
\end{equation}
where $H=\dot{a}/a$ is the Hubble parameter and a dot represents
differentiation with respect to $t$. To study the symmetries of
the minisuperspace under consideration, we need an effective
Lagrangian for the model whose variation with respect to its
dynamical variables yields the correct equations of motion.
Following \cite{7}, by considering the action described above as
representing a dynamical system in which the scale factor $a$ and
scalar curvature $R$ play the role of independent dynamical
variables, we can rewrite action (\ref{A}) as \cite{10}
\begin{equation}\label{E1}
{\cal S}=\int dt {\cal L}(a,\dot{a},R,\dot{R})=\int dt\left\{a^n
f(R)-\lambda\left[R-n(n-1)\frac{\dot{a}^2}{a^2}-2n\frac{\ddot{a}}{a}\right]\right\},
\end{equation}
where we introduce the definition of $R$ in terms of $a$ and its
derivatives as a constraint. This procedure allows us to remove
the second order derivatives from  action (\ref{E1}). The Lagrange
multiplier $\lambda$ can be obtained by variation with respect to
$R$, that is, $\lambda=a^n f'(R)$. Thus, we obtain the following
Lagrangian for the model \cite{7,10}
\begin{equation}\label{F}
{\cal L}(a,\dot{a},R,
\dot{R})=n(n-1)\dot{a}^2a^{n-2}f'+2n\dot{a}\dot{R}a^{n-1}f''+a^n(f'R-f).
\end{equation}
The momenta conjugate to variables $a$ and $R$ are
\begin{equation}\label{G}
p_a=\frac{\partial {\cal L}}{\partial
\dot{a}}=2n(n-1)\dot{a}a^{n-2}f'+2na^{n-1}\dot{R}f'',\end{equation}
\begin{equation}\label{H}
p_R=\frac{\partial {\cal L}}{\partial
\dot{R}}=2n\dot{a}a^{n-1}f''.
\end{equation}
The Hamiltonian corresponding to  Lagrangian (\ref{F}) can then be
written in terms of $a$, $\dot{a}$, $R$ and $\dot{R}$ as
\begin{equation}\label{I}
{\cal H}(a,\dot{a},R,
\dot{R})=n(n-1)\dot{a}^2a^{n-2}f'+2n\dot{a}\dot{R}a^{n-1}f''-a^n(f'R-f).
\end{equation}
Therefore, our cosmological setting is equivalent to a dynamical
system where the phase space is spanned by
$\left\{a,R,p_a,p_R\right\}$ with Lagrangian (\ref{F}) describing
the dynamics with respect to time $t$. Now, it is easy to see that
variation of  Lagrangian (\ref{F}) with respect to $R$ gives the
following well-known relation for the scalar curvature
\begin{eqnarray}
R=n(n-1)\frac{\dot{a}^2}{a^2}+2n\frac{\ddot{a}}{a},
\end{eqnarray}
while variation with respect to $a$ yields the field equation
(\ref{D}). Also, equation (\ref{E}) is nothing but the zero energy
condition ${\cal H}=0$ (Hamiltonian constraint). A quantum $f(R)$
cosmological model based on Hamiltonian (\ref{I}) can be found in
\cite{11}.

The setup for constructing the phase space and writing the
Lagrangian is now complete. In the next section we shall use the
above Lagrangian to find  the form of $f(R)$ that would admit a
Noether symmetry.
\section{The Noether symmetry}
As is well known, Noether symmetry approach is a powerful tool in
finding the solution to a given Lagrangian, including the one
presented above. In this approach, one is concerned with finding
the cyclic variables related to conserved quantities and
consequently reducing the dynamics of the system to a manageable
one. The investigation of Noether symmetry in the model presented
above is therefore the goal we shall pursue in this section. Here
our aim is to find the function $f(R)$ such that the corresponding
Lagrangian exhibits the desired symmetry. Following \cite{8}, we
define the Noether symmetry induced on the model by a vector field
$X$ on the tangent space $TQ=\left(a,R,\dot{a},\dot{R}\right)$ of
the configuration space $Q=\left(a,R\right)$ of Lagrangian
(\ref{F}) through
\begin{equation}\label{J}
X=\alpha \frac{\partial}{\partial a}+\beta
\frac{\partial}{\partial R}+\frac{d
\alpha}{dt}\frac{\partial}{\partial \dot{a}}+\frac{d
\beta}{dt}\frac{\partial}{\partial \dot{R}},
\end{equation}
such that the Lie derivative of the Lagrangian with respect to
this vector field vanishes
\begin{equation}\label{K}
L_X {\cal L}=0.
\end{equation}
In (\ref{J}), $\alpha$ and $\beta$ are functions of $a$ and $R$
and $\frac{d}{dt}$ represents the Lie derivative along the
dynamical vector field, that is,
\begin{equation}\label{L}
\frac{d}{dt}=\dot{a}\frac{\partial}{\partial
a}+\dot{R}\frac{\partial}{\partial R}.
\end{equation}
It is easy to find the constants of motion corresponding to such a
symmetry. Indeed, equation (\ref{K}) can be rewritten as
\begin{equation}\label{M}
L_X{\cal L}=\left(\alpha \frac{\partial {\cal L}}{\partial
a}+\frac{d\alpha}{dt}\frac{\partial {\cal L}}{\partial
\dot{a}}\right)+\left(\beta \frac{\partial {\cal L}}{\partial
R}+\frac{d\beta}{dt}\frac{\partial {\cal L}}{\partial
\dot{R}}\right)=0.
\end{equation}
Noting that $\frac{\partial {\cal
L}}{\partial q}=\frac{dp_q}{dt}$, we have
\begin{equation}\label{N}
\left(\alpha\frac{dp_a}{dt}+\frac{d\alpha}{dt}p_a\right)+\left(\beta\frac{dp_R}{dt}+\frac{d\beta}{dt}p_R\right)=0,
\end{equation}
which yields
\begin{equation}\label{O}
\frac{d}{dt}\left(\alpha p_a+\beta p_R\right)=0.
\end{equation}
Thus, the constants of motion are
\begin{equation}\label{P}
Q=\alpha p_a+\beta p_R.
\end{equation}
In order to obtain the functions $\alpha$ and $\beta$ we use
equation (\ref{M}). In general this equation gives a quadratic
polynomial in terms of $\dot{a}$ and $\dot{R}$ with coefficients
being partial derivatives of $\alpha$ and $\beta$ with respect to
the configuration variables $a$ and $R$. Thus, the resulting
expression is identically equal to zero if and only if these
coefficients are zero. This leads to a system of partial
differential equations for $\alpha$ and $\beta$. For Lagrangian
(\ref{F}), condition (\ref{M}) results in
\begin{eqnarray}\label{R}
&&
na^{n-1}(f'R-f)\alpha+a^nf''R\beta+2n\dot{R}^2\left(a^{n-1}f''\frac{\partial
\alpha}{\partial R}\right) \nonumber\\&&
+n\dot{a}^2\left[(n-1)(n-2)a^{n-3}f' \alpha+(n-1)a^{n-2}f'' \beta
+2(n-1)a^{n-2}f' \frac{\partial \alpha}{\partial a}+2a^{n-1}f''
\frac{\partial \beta}{\partial a}\right] \nonumber\\&&
+2n\dot{R}\dot{a}\left[(n-1)a^{n-2}f'' \alpha + a^{n-1} f''' \beta
+a^{n-1}f'' \frac{\partial \alpha}{\partial a}+(n-1)a^{n-2}f'
\frac{\partial \alpha}{\partial R}+a^{n-1}f'' \frac{\partial
\beta}{\partial R}\right] \nonumber \\ && =0,
\end{eqnarray}
which leads to the following system of equations
\begin{eqnarray}\label{T}
n(f'R-f)\alpha +a f''R\beta &=0,\\
f''\frac{\partial \alpha}{\partial R}&=0,\label{U}\\
(n-1)(n-2)f'\alpha + (n-1)a f''\beta +2(n-1)af'\frac{\partial
\alpha}{\partial a}+2a^2f''\frac{\partial \beta}{\partial
a}&=0,\label{V}\\
 (n-1)f''\alpha+af'''\beta+af''\frac{\partial
\alpha}{\partial a}+(n-1)f'\frac{\partial \alpha}{\partial R}+a
f''\frac{\partial \beta}{\partial R}&=0. \label{W}
\end{eqnarray}
Equation (\ref{U}) gives $f''=0$ or $\frac{\partial
\alpha}{\partial R}=0$. In the case where $f''=0$, we obtain
\begin{eqnarray}
f(R)=c_1R+c_2.
\end{eqnarray}
Substituting this result into equation (\ref{T}) we get $c_2=0$
and recover the usual Einstein-Hilbert gravity without
cosmological constant for which $f(R)=R$. Also, since equation
(\ref{W}) results in $\frac{\partial \alpha}{\partial R}=0$, from
equation (\ref{V}) we obtain the following differential equation
for $\alpha (a)$
\begin{equation}\label{X}
(n-2)\alpha(a)+2a\frac{d\alpha}{da}=0,
\end{equation}
with solution
\begin{equation}\label{Y}
\alpha(a)=a^{-\frac{n-2}{2}}.
\end{equation}
Therefore, for any arbitrary function $\beta(a,R)$, the following
vector field represents a Noether symmetry for the
minisuperspace of the flat FRW cosmology in Einstein-Hilbert
gravity
\begin{equation}\label{Z}
X=a^{-\frac{n-2}{2}}\frac{\partial}{\partial
a}-\frac{n-2}{2}\dot{a}a^{-\frac{n}{2}}\frac{\partial}{\partial
\dot{a}}+\beta \frac{\partial}{\partial
R}+\dot{\beta}\frac{\partial}{\partial \dot{R}}.
\end{equation}
Since from (\ref{H}) we have $p_R=0$, the corresponding constant
of motion is $Q=\alpha p_a$, which yields
\begin{equation}\label{AB}
a^{-\frac{n-2}{2}}\dot{a}a^{n-2}=\mbox{const.} \Rightarrow
a(t)\sim t^{2/n}.
\end{equation}
Although, this solution for the scale factor satisfies the field
equation (\ref{D}), it does not satisfy the Hamiltonian constraint
(\ref{E}). This is not surprising since it is well known that the
flat FRW cosmology with Einstein-Hilbert action (without the
cosmological constant) has no vacuum solution. Thus, we remove the
case $f''=0$ from our consideration. When $f'' \neq 0$, from
equation (\ref{U}) we obtain
\begin{eqnarray}
\frac{\partial \alpha}{\partial R}=0.
\end{eqnarray}
Under this condition equation (\ref{T}) results in
\begin{equation}\label{AC}
\alpha(a)=\frac{af''R}{n(f-f'R)}\beta(a,R).
\end{equation}
If one uses this expression in equation (\ref{V}) to eliminate
$\alpha(a)$, one obtains
\begin{equation}\label{AD}
\frac{\partial \beta}{\partial a}=\frac{n(n-1)f}{2a(f'R-nf)}\beta
(a,R).
\end{equation}
To solve this equation we assume that the function $\beta (a,R)$
can be written in the form $\beta (a,R)=A(a)B(R)$, where $A$ and
$B$ are functions only of $a$ and $R$ respectively. Substituting
this ansatz for $\beta (a,R)$ into equation (\ref{AD}), we obtain
\begin{equation}\label{AE}
\frac{2a}{A}\frac{dA}{da}=\frac{n(n-1)f}{f'R-nf}.
\end{equation}
Since the left-hand side of this equation is a function of $a$
only while the right-hand side is a function of $R$, we should
have
\begin{equation}\label{AF}
\frac{n(n-1)f}{f'R-nf}=c=\mbox{Const.},
\end{equation}
which results in
\begin{equation}\label{AG}
f(R)=R^{\frac{n(n+c-1)}{c}},
\end{equation}
where the constant $c\neq -n$ is required to have $f''\neq 0$. On
the other hand equation (\ref{AE}), with its right hand-side equal
to $c$, has the solution
\begin{equation}\label{AH}
\frac{2a}{a}\frac{dA}{da}=c \Rightarrow
A(a)=a^{c/2}.
\end{equation}
Now, using (\ref{AG}) and $\beta(a,R)=a^{c/2}B(R)$ in equation
(\ref{AC}), we find
\begin{equation}\label{AI}
\alpha(a)=-\frac{n+c-1}{c}R^{-1}a^{\frac{c}{2}+1}B(R).
\end{equation}
Since $\alpha(a)$ should be a function of $a$ only, from the above
expression for $\alpha(a)$ we conclude that $B(R)=R$ and thus
obtain
\begin{equation}\label{AJ}
\beta(a,R)=Ra^{c/2},\hspace{.5cm}\alpha(a)=-\frac{n+c-1}{c}a^{\frac{c}{2}+1}.
\end{equation}
To determine the constant $c$, we note that relations (\ref{AG})
and (\ref{AJ}) should satisfy equation (\ref{W}). Thus, by
substituting these results into equation (\ref{W}) we find
\begin{equation}\label{AK}
c=-1-n,
\end{equation}
which completes our solutions as
\begin{equation}\label{AL}
\alpha(a)=-\frac{2}{n+1}a^{-\frac{n-1}{2}},\hspace{.5cm}\beta(a,R)=Ra^{-\frac{n+1}{2}},
\end{equation}
and
\begin{equation}\label{AM}
f(R)=R^{\frac{2n}{n+1}}.
\end{equation}
Therefore, in the context of $f(R)$ cosmology, a flat FRW metric
has Noether symmetry if the corresponding action is given by
equation (\ref{AM}) and the Noether symmetry is generated by the
following vector field
\begin{eqnarray}\label{AN}
X&=&-\frac{2}{n+1}a^{-\frac{n-1}{2}}\frac{\partial}{\partial
a}+Ra^{-\frac{n+1}{2}}\frac{\partial}{\partial
R}+\frac{n-1}{n+1}\dot{a}a^{-\frac{n+1}{2}}\frac{\partial}{\partial
\dot{a}}\nonumber
\\&&+\left(\dot{R}a^{-\frac{n+1}{2}}-\frac{n+1}{2}R\dot{a}a^{-\frac{n+3}{2}}\right)\frac{\partial}{\partial
\dot{R}}.
\end{eqnarray}
To obtain the corresponding cosmology resulting from this type of
$f(R)$, we note that the existence of Noether symmetry (\ref{AN})
implies the existence of a constant of motion $Q=\alpha p_a+\beta
p_R.$ Hence, using equation (\ref{G}) and (\ref{H}) we have
\begin{equation}\label{AO}
Q=-\frac{4n^2(n-1)}{(n+1)^2}\dot{a}a^{\frac{n-3}{2}}R^{\frac{n-1}{n+1}}-\frac{8n^2(n-1)}{(n+1)^3}a^{\frac{n-1}{2}}\dot{R}
R^{-\frac{2}{n+1}},
\end{equation}
which may be rewritten in the form
\begin{equation}\label{AP}
Q=-\frac{8n^2}{(n+1)^2}\frac{d}{dt}\left(a^{\frac{n-1}{2}}R^{\frac{n-1}{n+1}}\right),
\end{equation}
and can be immediately integrated with the result, assuming
$a(t=0)=0$
\begin{equation}\label{AQ}
a^{\frac{n-1}{2}}R^{\frac{n-1}{n+1}}=-\frac{(n+1)^2}{8n^2}Qt.
\end{equation}
On the other hand, the Hamiltonian constraint ${\cal H}=0$ with
${\cal H}$ given by (\ref{I}) gives
\begin{equation}\label{AR}
2n^2\dot{a}^2R^{-1}+\frac{4n^2}{n+1}\dot{a}\dot{R}aR^{-2}=a^2.
\end{equation}
To obtain the scale factor from the above relation, note that
equation (\ref{AQ}) yields
\begin{equation}\label{AS}
\dot{R}=\left[-\frac{(n+1)^2}{8n^2}Q\right]^{\frac{n+1}{n-1}}\left(\frac{n+1}{n-1}t^{\frac{2}{n-1}}a^{-\frac{n+1}{2}}-
\frac{n+1}{2}t^{\frac{n+1}{n-1}}\dot{a}a^{-\frac{n+3}{2}}\right),
\end{equation}
which, upon substitution into relation (\ref{AR}), gives
\begin{equation}\label{AT}
\frac{4n^2}{n-1}\left[-\frac{(n+1)^2}{8n^2}Q\right]^{-\frac{n+1}{n-1}}\dot{a}a^{\frac{n-1}{2}}=t^{\frac{2n}{n-1}},
\end{equation}
where, after integration we obtain
\begin{equation}\label{AU}
a(t) \sim t^{\frac{2(3n-1)}{n^2-1}}.
\end{equation}
Therefore, in the context of our $f(R)=R^{\frac{2n}{n+1}}$
gravity, the universe evolves with a power law expansion (note
that for any $n>1$ the power of the scale factor is positive). It
is remarkable from (\ref{AU}) that the condition under which the
universe would accelerate is $\frac{2(3n-1)}{n^2-1}>1$, that is,
$n\leq 5$. This means that models with spatial dimension $n\leq 5$
obey an accelerated power law expansion while for $n>5$ a
decelerated expansion occurs.
\section{The equation of state parameter}
One of the advantages of $f(R)$ theories of gravity is that they
can describe and provide gravitational alternative for dark
energy. Indeed, the equations of motion resulting from $f(R)$
gravity admit such solutions which predict the same accelerated
expansion as those resulting from the usual Einstein-Hilbert
gravity with dark energy. In other words, in the context of $f(R)$
cosmology one can describe the accelerated expansion of the
universe without the necessity to introduce the exotic fluid with
a negative equation of state (EoS) parameter $w$. To introduce an
effective EoS parameter in our model, one may compare the field
equations (\ref{D}) and (\ref{E}) with the usual field equations
of a flat FRW universe filled with a perfect fluid with EoS
$P=w\rho$, that is, with $3H^2\sim \rho$ and $-2\dot{H}-3H^2 \sim
P$. This comparison shows that in our model one may introduce the
effective energy density and pressure as follows
\begin{equation}\label{AW}
\rho=\frac{1}{2}\left(Rf'-f\right)-nH\dot{R}f'',
\end{equation}
\begin{equation}\label{AX}
P=f'''\dot{R}^2+(n-1)H\dot{R}f''+\ddot{R}f''+\frac{1}{2}\left(f-Rf'\right).
\end{equation}
Therefore, we can define the effective EoS parameter $w_{eff}$ as
\begin{equation}\label{AY}
w_{eff}=\frac{P}{\rho}=\frac{f'''\dot{R}^2+(n-1)H\dot{R}f''+\ddot{R}f''+\frac{1}{2}\left(f-Rf'\right)}
{\frac{1}{2}\left(Rf'-f\right)-nH\dot{R}f''}.
\end{equation}
To transform the above expression to a more manageable form,
consider the quantity ${\cal F}=f'(R)$ \cite{7}. In terms of ${\cal
F}$, equation (\ref{AY}) takes the form
\begin{equation}\label{AZ}
w_{eff}=-1+\frac{2}{n(n-1)}\frac{\ddot{{\cal F}}-H\dot{{\cal
F}}}{{\cal F}H^2}.
\end{equation}
For the problem at hand, taking into account that
\begin{eqnarray}
f'(R)=\frac{2n}{n+1}R^{\frac{n-1}{n+1}},
\end{eqnarray}
and with the help of equation (\ref{AQ}) we obtain
\begin{equation}\label{BA}
{\cal F}=-\frac{n+1}{4n}Qt^{-2\frac{n-1}{n+1}}.
\end{equation}
Thus, substituting (\ref{BA}) and $H=\dot{a}/a$ from (\ref{AU}) we
are led to the following EoS parameter
\begin{equation}\label{BC}
w_{eff}=-1+\frac{(n-1)(n+1)}{n(3n-1)}.
\end{equation}
Now, it is easy to see that $-1<w_{eff}<0$, which is the
characteristic of one type of dark energy, the so-called
quintessence. Note that when $w_{eff}$ is less than $-1$, the EoS
describes another type of dark energy known as phantom. As is clear
from equation (\ref{BC}), the effective EoS parameter is always
greater than $-1$ and thus in the model under consideration here,
the phantom phase cannot be accounted for.
\section{Conclusions}
In this letter we have studied a generic $f(R)$ cosmological model
with an eye to Noether symmetry. For the background geometry, we
have considered a flat ($n$+1)-dimensional FRW metric and derived the
general equations of motion in this background. The phase space was
then  constructed by taking the scale factor $a$ and Ricci scalar
$R$ as the independent dynamical variables. The Lagrangian of the
model in the configuration space spanned by $\left\{a,R\right\}$ is
so constructed such that its variation with respect to these
dynamical variables yields the correct field equations. The
existence of Noether symmetry implies that the Lie derivative of
this Lagrangian with respect to the infinitesimal generator of the
desired symmetry vanishes. By applying this condition to the
Lagrangian of the model, we have obtained the explicit form of the
corresponding $f(R)$ function. We have shown that this form of
$f(R)$ results in a power law expansion for the scale factor of the
universe and the expansion is accelerating in the case of $n\leq 5$.
We have also presented an effective EoS parameter for our $f(R)$
cosmology model. Our analysis shows that the EoS parameter is always
greater than $-1$ and thus the so-called phantom dark energy cannot
be described in this kind of modified gravity. On the other hand,
the EoS parameter is restricted to the interval $-1<w_{eff}<0$.
Since this interval is characteristic of the quintessence dark
energy, our $f(R)$ model can provide a natural gravitational
alternative for this kind of dark energy without the necessity to
introduce an exotic fluid with a negative EoS parameter.
 \vspace{5mm}\newline \noindent {\bf
Acknowledgement}\vspace{2mm}\noindent\newline The author would like
to thank H.R. Sepangi and N. Khosravi for a careful reading of the
manuscript and useful comments.

\end{document}